\documentclass[a4paper,11pt]{article}
\usepackage{aaskaiid}
\usepackage{graphicx}
\usepackage{orcidlink}

\title{Application of Bayesian Statistical Tools to SKA Telescopes Polarization Surveys to Study Magnetization of the Large-scale Structure of the Universe} 
\ShortTitle{Application of Bayesian statistical tools to SKA telescopes polarization surveys}

\author[1,2]{Valentina Vacca\orcidlink{0000-0003-1997-0771}}
\ShortName{Vacca et al.} 
\author[3]{Sebastian Hutschenreuter}
\author[1,2,4]{Andrea Cabriolu\orcidlink{0009-0004-6735-4062}}
\author[2,5]{Torsten A.\,En{\ss}lin}
\author[2,6]{Philipp Frank}
\author[2,7]{Jakob Roth\orcidlink{0000-0002-8873-8215}}
\author[2]{Martin Reinecke}
\author[8]{Jens Jasche}
\author[9]{Florent Leclercq\orcidlink{0000-0002-9339-1404}}
\author[10]{Ettore Carretti}
\author[10]{Luigina Feretti}
\author[11]{Chiara Ferrari}
\author[1]{Federica Govoni}
\author[12]{Cathy Horellou\orcidlink{0000-0002-3533-8584}}
\author[13]{Melanie Johnston-Hollitt\orcidlink{0000-0003-2756-8301}}
\author[1]{Francesca Loi}
\author[1]{Matteo Murgia}
\author[14]{Shane O'Sullivan\orcidlink{0000-0002-3968-3051}}
\author[11]{Rosita Paladino}
\author[15,16]{Tessa Vernstrom}
\author[4]{Gianni Fenu}

\affiliation[1]{INAF-Osservatorio Astronomico di Cagliari, Via della Scienza 5, I-09047 Selargius (CA), Italy}
\affiliation[2]{Max Planck Institute for Astrophysics, Karl-Schwarzschildstr. 1, 85741 Garching, Germany}
\emailAdd{valentina.vacca@inaf.it}
\affiliation[3]{University of Vienna, Department of Astrophysics, T\"urkenschanzstra{\ss}e 17, 1180 Vienna, Austria}
\affiliation[4]{Department of Mathematics and Computer Science, University of Cagliari, Via Ospedale 72, Cagliari, 09121, Italy}
\affiliation[5]{Deutsches Zentrum für Astrophysik, Postplatz 1, 02826 Görlitz, Germany}
\affiliation[6]{Kavli Institute for Particle Astrophysics \& Cosmology, Stanford University, Stanford, CA 94305, USA}
\affiliation[7]{Technische Universität M\"unchen (TUM), Boltzmannstr. 3, 85748 Garching,
Germany}
\affiliation[8]{The Oskar Klein Centre, Department of Physics, Stockholm University,
AlbaNova University Centre, SE 106 91 Stockholm, Sweden}
\affiliation[9]{CNRS \& Sorbonne Université, UMR 7095, Institut d’Astrophysique de Paris, 98 bis boulevard Arago, F-75014 Paris,
France}
\affiliation[10]{INAF - Istituto di Radioastronomia, Via P. Gobetti 101, 40129 Bologna, Italy}
\affiliation[11]{Université Côte d'Azur, OCA, CNRS, Laboratoire Lagrange, Boulevard de l'Observatoire, CS 34229, 06304 Nice Cedex 4, France}
\affiliation[12]{Chalmers University of Technology, Department of Space, Earth and Environment, Onsala Space Observatory, 439 92 Onsala, Sweden}
\affiliation[13]{Curtin Institute for Data Science, Curtin University, Perth, WA, Australia}
\affiliation[14]{Departamento de F{\'i}sica de la Tierra y Astrof{\'i}sica \& IPARCOS-UCM, Universidad Complutense de Madrid, 28040 Madrid, Spain}
\affiliation[15]{CSIRO Space and Astronomy, PO Box 1130, Bentley, WA 6102, Australia}
\affiliation[16]{ICRAR, The University of Western Australia, 35 Stirling Hw, 6009 Crawley, Australia}

\abstract{Understanding cosmological magnetic fields requires a detailed knowledge of magnetism in the different environments of the large-scale structure of the Universe. Magnetic fields are well known to inhabit galaxy clusters, and recently their presence has been detected between galaxy clusters, along filaments extending up to 10-15 Mpc.
Beyond that, there is limited
information on the existence
of magnetic fields in sheets and voids of the cosmic web. We propose a Bayesian statistical approach to study magnetic fields on large scales through observations of the Faraday rotation effect in large samples of polarized point-like background radio sources. We present the expectations to detect magnetization in environments of the large-scale structure with the SKA-Mid polarization survey planned by the SKAO Magnetism Science Working Group and with SKA-Low with AA4 telescopes, and discuss the required level of accuracy on the redshifts of the host galaxies for such a study.
We find that about 50,000 mid-frequency Faraday rotation measurements complemented by high-precision redshifts are needed to constrain magnetization of dense environments as galaxy clusters. Investigation of magnetization in weakly-magnetized low-density enviroments, as filaments, will remain challenging, but low frequencies radio observations and spectroscopic redhifts for at least 17,000 will allow us to put first constraints.}

\begin{document}
\maketitle

\section{Introduction}
Understanding the origin and evolution of large-scale magnetic fields in the Universe is a key science goal in the use of present and future-generation radio telescopes. Despite magnetic fields 
been observed up to galaxy cluster scales and in filaments between clusters through synchrotron emission (e.g., \citealt{Govoni2019}, \citealt{Vernstrom2023}), firm constraints on magneto-genesis scenarios have been not placed yet.
Gamma-rays observations of blazars put a lower limit on magnetic fields in cosmic voids of a few fG \citep[see e.g. ][for a recent work]{Tjemsland2024}, favoring a primordial magneto-genesis scenario as suggested also by radio observations \citep{Carretti2025}.
A powerful tool to detect and characterize extragalactic magnetic fields is 
Faraday rotation of the plane of polarization of radio emission of  background linearly polarized radio galaxies \citep[e.g.,][]{Govoni2004}. 
Generally speaking, the Faraday rotation effect can be described through the Faraday depth \citep{Burn1966,Kronberg2008},
\begin{equation}
\phi (z)\propto\int_0^z\frac{d\mathrm{l}}{d\mathrm{z^{\prime}}}\frac{n_{\rm e}(z^{\prime})B_{\rm l}(z^{\prime})}{(1+z^{\prime})^2}d\mathrm{z^{\prime}},
\end{equation}
where $n_{\rm e}$ is the thermal electron gas density, $B_{\rm l}$ the magnetic field component along the line of sight, $z$ the redshift of the source and $d\mathrm{l}/d\mathrm{z^{\prime}}$ is the comoving path increment per unit redshift.
The polarization angle is rotated by an angle $\phi\lambda^2$ while the emission crosses the magneto-ionic media between the emitting radio source and the observer. These media are our own Galaxy, galaxy clusters, filaments, voids, sheets, other intervening galaxies and the emitting radio source itself \citep[e.g., ][]{Vacca2015}. 

Polarimetric surveys provide measurements of the Faraday rotation effect for a collection of extragalactic radio sources, the so-called rotation measure (RM) grids, that carry information about magnetization in this variety of enviroments, i.e. all the media traversed by the polarized signal on its way from the source to the observer, and can therefore be used to study magnetization from small-scale systems, as stars, to the largest scales, as galaxy clusters and the cosmic web. 
Because the different line of sight to sources probes different combination of Faraday rotating environments, identifying the signature of the weak magnetic fields in filaments and voids becomes thinkable, while remaining extremely challenging. The signature of extragalactic magnetism is indeed often buried in the observing noise and  sophisticated statistical tools are required for their detection. The first step in this direction is a reliable and accurate reconstruction of the Galactic Faraday rotation. An image of the Galactic Faraday rotation using most of the data presently available has been recently produced by \cite{Hutschenreuter2022}, with a renewed and sophisticated Bayesian approach based on Information Field Theory \citep{Ensslin2009}. In \cite{Vacca2026}, we extended this algorithm in order to simultaneously statistically disentangle the Galactic and the extragalactic contributions, properly taking into account the observing noise, exploiting previous work published in \cite{Vacca2015,Vacca2016} built on \cite{Oppermann2015} results. 
A precise estimate of the distance of each source in the RM catalogue together with information about the component of the intervening large-scale structure will allow us to indentify the extragalactic Faraday
contributions of galaxy clusters, filaments, sheets and voids. To this end, the availability of complementary \emph{spectroscopic} redshifts is crucial. 

The Magnetism SKAO Science Working Group identified as a top priority a polarimetric survey with SKA-Mid over 30,000 square degrees in band\,2 (950-1760\,MHz) at 2$^{\prime\prime}$ down to a sensitivity of 4\,$\mu$Jy corresponding to an observing time of about 15\,min per pointing \citep{Heald2020}. In this paper, we present predictions obtained by applying our algorithm to synthetic data from an SKA-Mid survey with AA4 telescopes, 
with specifications similar to the real ones. 
Observations at mid-frequencies will allow us to investigate magnetic fields in dense and highly-magnetized systems, as galaxy clusters, and characterize their properties, e.g., their dependence on physical parameters like the redshift of the emitting radio source. Moreover, recently, LOw Frequency ARray (LOFAR) observations revealed that low frequency observations are particularly suitable to detect and characterize magnetic fields in low-density weakly magnetized environments, such as filaments of the cosmic web \citep{Carretti2022}. These frequencies indeed enable high-precision Faraday rotation measurements, crucial to detect the faint signature of magnetization in filaments. At these frequencies, the signal is depolarized as it passes through environments with strong magnetic fields and high thermal gas densities. Consequently, the polarized signal that can be detected at meter-wavelengths has been mainly rotated by low-density weakly magnetized structures. The LOFAR Two-metre Sky Survey (LoTSS) delivered Faraday rotation measurements for a catalogue of 2461 extragalactic radio sources over about 6000\,deg$^2$ of the sky, translating in a number density of polarized sources of about 0.43\,deg$^{-2}$ \citep{OSullivan2023}. Findings by \cite{Carretti2023} and \cite{Carretti2025} based on these data suggest that they are dominated by the Faraday rotation imprinted on the polarized signal while crossing filaments of the cosmic web and obtained indication favouring primordial magneto-genesis scenarios. 

In order to shed light on the origin and evolution of cosmic magnetism, a survey spanning the frequency range from low- to mid-frequencies would be extremely precious. The availability of a dense grid of measurements at mid-frequencies, would be important for an accurate removal of the Galactic foreground, crucial not only to shed light on magnetic field history through mid-frequencies but also to constrain the magneto-genesis with the aid of low frequency data. For these reasons, in this work, we also show predictions by applying our algorithm to synthetic data from an SKA-Low survey with AA4 telescopes. 

In the following, we briefly describe our algorithm in \S\,\ref{algorithm}, we present and discuss the expectations for SKA-Mid and SKA-Low AA4 telescopes in \S\,\ref{expectations}, and we draw our conclusions in \S\,\ref{conclusions}.

\section{Description of the algorithm}
\label{algorithm}
We expanded the Bayesian algorithm of \cite{Hutschenreuter2022} in order to apply it to modern Faraday rotation catalogues to detect and characterize extragalactic magnetic fields.
The data of interest are Faraday rotation measurements in the direction of $N$ polarized point-like background radio sources, denoted $d = \{d_i\}^{\rm N}_{\rm i=1} = d_{\rm 1}, d_{\rm 2}, ..., d_{\rm N}$.

\cite{Hutschenreuter2022} model the overall Faraday rotation as
\begin{equation}
d_i = \phi_{\rm gal,i} + \tilde{n_{\rm i}},
\end{equation}
where the extragalactic contribution, if present, is absorbed in the increased noise term $\tilde{n_{\rm i}}$:
\begin{equation}
\tilde{n_{\rm i}}=\phi_{\rm eg,i} + n_{\rm i}.
\end{equation}
The noise covariance matrix $\tilde{N}$ consists of the measurement uncertainties $\sigma_{\rm n}$ increased by a correction factor, $\eta$, 
\begin{equation}
    \tilde{N}=\mathrm{diag}(\eta\sigma_{\rm n}^2).
    \label{galvar}
\end{equation}
The $\eta$-factors are computed for each data point individually through inference from the data and include the extragalactic contribution and possible corrections for unreliable estimates of the observing noise.
For the fraction of sources with spectroscopic redshift and auxiliary radio information (radio luminosity),
we model the Faraday rotation as a sum of a Galactic, an extragalactic term and the observing noise,
\begin{equation}
d_i = \phi_{\rm gal,i} + \phi_{\rm eg,i} + n_{\rm i}.
\end{equation}

Expanding on \cite{Hutschenreuter2022}, the algorithm described here is capable to distinguish between the Galactic, the extragalactic term and the noise contribution, by exploiting the angular correlation of the Galactic component, and the fact that the extragalactic and noise contributions can be assumed to be uncorrelated on the sky. This extended algorithm 
operates as follows: a generative revised stochastic model for the data is set up. This includes as model components physical quantities of interest, like the amount of expected Faraday rotation dispersion per length associated with the overall extragalactic large-scale structure between the source and the observer or, alternatively, a certain cosmic environment (galaxy cluster, filament, ...). Then the parameters of the generative model are inferred conditional to the observed data. The result is a Bayesian posterior distribution of all model parameters. As field like quantities have to be inferred simultaneously, the Galactic Faraday sky, the inference problem is very high dimensional. To deal with this in finite computational time, approximative variational inference methods are used \citep[i.e. geometrical Variational Inference, ][]{Frank2021}. These technical advancements represent a significant improvement with respect to the work of \cite{Vacca2016}. 

The functioning of the algorithm relies on two catalogues. A catalogue contains data points used to infer the Galactic contribution only. For these data, the noise distribution is assumed to be a zero-mean Gaussian with variance given by Eq.\,\ref{galvar}. The remaining data points, for which complementary information is available (as, e.g., spectroscopic redshifts), are assumed to follow a Gaussian noise distribution with zero-mean and standard deviation given by the observing uncertainty. For these data points, the distribution of extragalactic Faraday rotation values is modelled as a zero-mean Gaussian, in agreement with previous work by \cite{Oppermann2015,Schnitzeler2010,Clarke2004}. 
The distribution of extragalactic Faraday rotation likely is non-Gaussian, as radio observations suggest: Faraday rotation measures up to thousands of rad/m$^2$ have been observed in cool core galaxy clusters that are not explainable with a Gaussian distribution characterized by a standard deviation of a few rad/m$^2$ \citep[e.g.,][]{Oppermann2015}. In the following, a Gaussian prior for the extra-galactic contribution is assumed per RM data point, but we note that this is not in conflict with the overall distribution of extragalactic RM-values being non-Gaussian, as these distributions live on entirely different domains.

We parameterize the variance of this Gaussian as a sum of contributions intrinsic to the emitting source $\sigma_{\rm int, i}^2$ and associated with the large-scale structure environment between the source and the observer $\sigma_{\rm env, i}^2$. Overall, we model it as 
\begin{equation}
\begin{split}
\sigma_{\rm eg, i}^2 = \sigma_{\rm int, i}^2 + \sigma_{\rm env, i}^2=\\
&
=\left(\frac{L_{i}}{L_0}\right)^{\chi_{\rm lum}}\frac{e^{\chi_{\rm int, 0}}}{(1+z_{i})^4} +e^{\chi_{\rm env, 0}}\frac{1}{D_0}\int_0^{z_{i}}\frac{c}{H(z)}(1+z)^{4+\chi_{\rm red}}d\mathrm{z},
\end{split}
\label{egcontr}
\end{equation}
where $L_i$ and $z_i$ are, respectively, the radio luminosity and the redshift of the emitting radio source, $L_0$ and $D_0$ are two normalization constants, $H(z)$ is the Hubble parameter, $c$ is the speed of light \citep[see][for details about the derivation]{Vacca2016}. The parameters $\chi_{\rm lum}$ and $\chi_{\rm red}$ describe the luminosity and redshift dependence, while $e^{\chi_{\rm int, 0}}$ and $e^{\chi_{\rm env, 0}}$ are the pure (with no further dependencies on luminosity and/or redshift) Faraday rotation variances, respectively of the emitting radio source and of the remaining extragalactic environments along the line of sight. In total, the extragalactic model has four parameters: $\Theta=\{\chi_{\rm int,0}, \chi_{\rm env,0}, \chi_{\rm lum}, \chi_{\rm red}\}$.
We note that this model does not take into account all possible correlations: for example, we do not include in the environmental Faraday rotation variance any dependence on luminosity. In this work, we use a simple modeling since our aim is to show the potential of our algorithm for the study of extragalactic magnetization, when catalogues as those expected from SKA-Mid and SKA-Low observations are used. 

A more in depth description of the algorithm is presented by \cite{Vacca2026}.

\section{Expectations with the SKA AA4 telescopes}
\label{expectations}

\subsection{Polarization source counts}
\label{sourcecounts}

At low frequencies, \cite{OSullivan2026}, in this book, predict that the cumulative number $N$ of polarized sources per square
degree scales as a function of the polarized flux density $P$ as
\begin{equation}
    N(>P) \sim 5 \left(\frac{P}{100\,\mu Jy}\right)^{-0.75}~~~ {\rm deg^{-2}},
\end{equation}
by extrapolating information between $\sim$0.1 and 10\,mJy based on T-RECS total intensity counts at 150\,MHz \citep{Lin2024} and deep field polarimetric LOFAR counts by \cite{Piras2024}. 
O'Sullivan et al.\ compute that, for a survey over 10,000\,deg$^2$ of the sky with 8\,h pointings, a conservative 80\,$\mu$Jy/beam sensitivity can be assumed, corresponding to about 60,000 polarized sources. At these frequencies, the Faraday rotation uncertainty is about 0.06\,rad/m$^2$ and is dominated by the ionosphere, as also shown for LOFAR data \citep{OSullivan2023}.

At mid-frequencies, 
the density of polarized sources can be derived from the work of \cite{Stil2014}, see Table\,1 in that paper, and from the
polarization number counts per square degree derived by \cite{Rudnick2014},
\begin{equation}
        N(>P) \sim 45 \left(\frac{P}{30\,\mu Jy}\right)^{-0.6}~~~ {\rm deg^{-2}}.
\end{equation}
By exploiting these findings, \cite{Heald2020} derive a density of polarized sources of 60-90\,deg$^{-2}$ at the sensitivity of the all-sky survey defined by the SKAO Magnetism Science Working Group ($\sim$4\,$\mu$Jy/beam per Stokes parameter over the full bandwidth and for a spatial resolution of 2$^{\prime\prime}$).
Considering the most conservative estimates, these number counts translate into an overall number of polarized sources of about 1\,million for the full southern sky.

\subsection{Generation of synthetic catalogues}
In order to assess the performances of data from SKA AA4 telescopes, we produced synthetic catalogues based on the rotation measure catalogue of \cite{VanEck2023}, version number 1.2.0, that includes most of the Faraday measurements available to June 2024.
We generated synthetic values for the Faraday rotation, its uncertainty, the redshift and the luminosity in Stokes I. The uncertainty in redshift, luminosity and coordinates has been assumed to be negligible. The spatial distribution of the synthetic sources reflects the coverage of the surveys considered. 

\begin{figure}[h]
    \centering
    	\includegraphics[width=0.7\columnwidth]{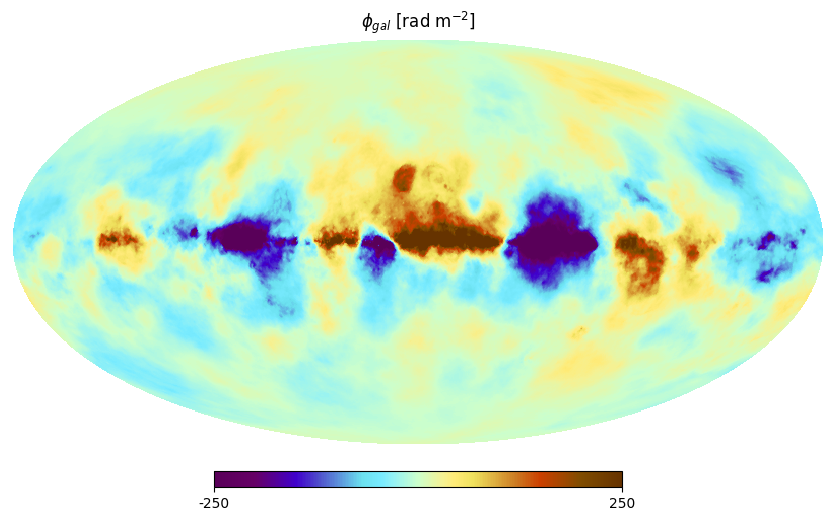}\\
   
    \caption{Synthetic Galactic Faraday sky. }
    \label{fig:galphi}
\end{figure}

\begin{figure}[h!]
    \centering
	\includegraphics[width=0.75\columnwidth]{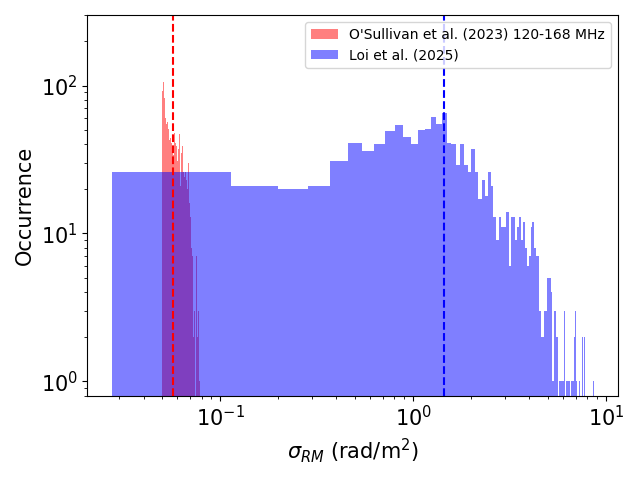}
	\includegraphics[width=0.75\columnwidth]{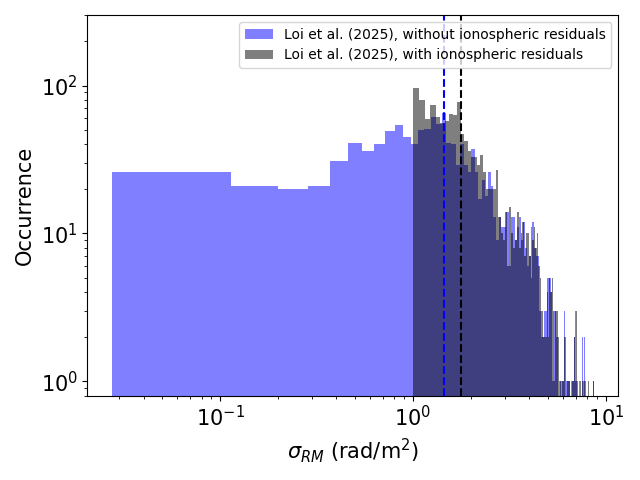}
    \caption{\emph{Top panel:} histogram of the uncertainties in the Faraday rotation measurements from the LoTSS DR2 RM catalogue \citep{OSullivan2023} in red colors and from the MeerKAT data \citep{Loi2025} in blue colors. \emph{Bottom panel:} histogram of the uncertainties in the Faraday rotation measurements from the MeerKAT data \citep{Loi2025} at 0-9-1.4\,GHz in blue colors with no ionospheric corrections and in black colors including an ionospheric correction residual of 1\,rad/m$^2$. The dashed vertical lines in both panels indicate the median of each distribution. The median of the uncertainties in the LOTSS DR2 RM catalogue is $\approx$0.06\,rad/m$^2$, in the MeerKAT catalogue with no correction for the ionospheric Faraday rotation is $\approx$1.5\,rad/m$^2$ and including ionospheric correction residuals is $\approx$1.8\,rad/m$^2$. }
    \label{fig:sigma_rm}
\end{figure}

\begin{figure}[h!]
    \centering
    	\includegraphics[width=\columnwidth]{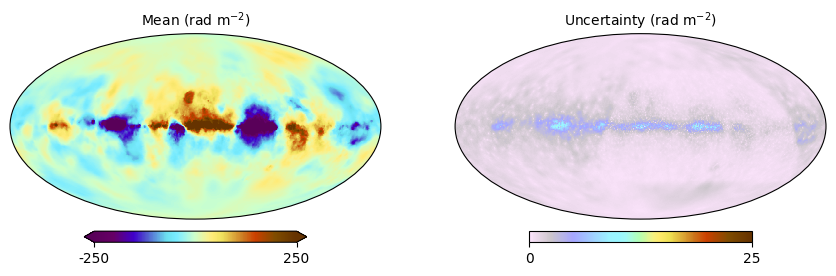}\\
\includegraphics[width=\columnwidth]{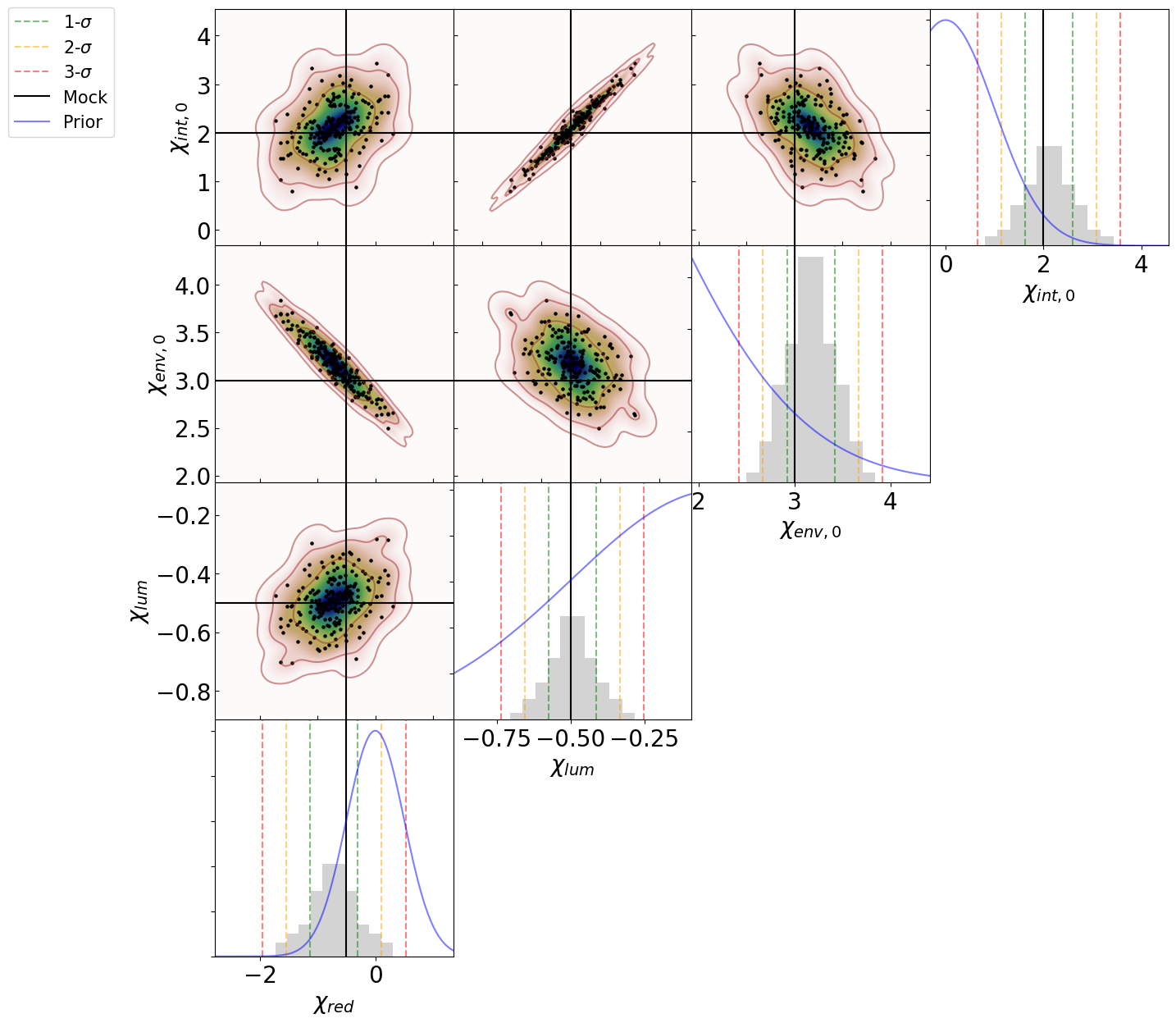}
    \caption{Results obtained assuming an extragalactic Faraday rotation standard deviation of $\approx$10\,rad/m$^2$ and using a rotation measure catalogue based on observations at mid-frequencies. \emph{Top panels:} Reconstructed mean and uncertainty of the Galactic Faraday sky. \emph{Bottom panels} 2-dimensional marginalization (in colours) and 1-dimensional marginalization (grey histograms) of the posterior distribution of the extragalactic parameters. The black line represents the ground truth value of the parameter used in the synthetic data. Brown contours represent the 1-, 2-, and 3-$\sigma$ of the 2-dimensional marginalizations of the posterior distribution. Dashed green, orange and red lines, the 1-, 2-, and 3-$\sigma$ of the 1-dimensional posterior distribution. The blue line shows the prior distribution. Black dots are random samples from the posterior distribution. Mirror images of the samples around the mean have also been included.}
    \label{fig:skamid}
\end{figure}

\begin{figure}[h]
    \centering
	\vspace{0.5cm}
    \includegraphics[width=\columnwidth]{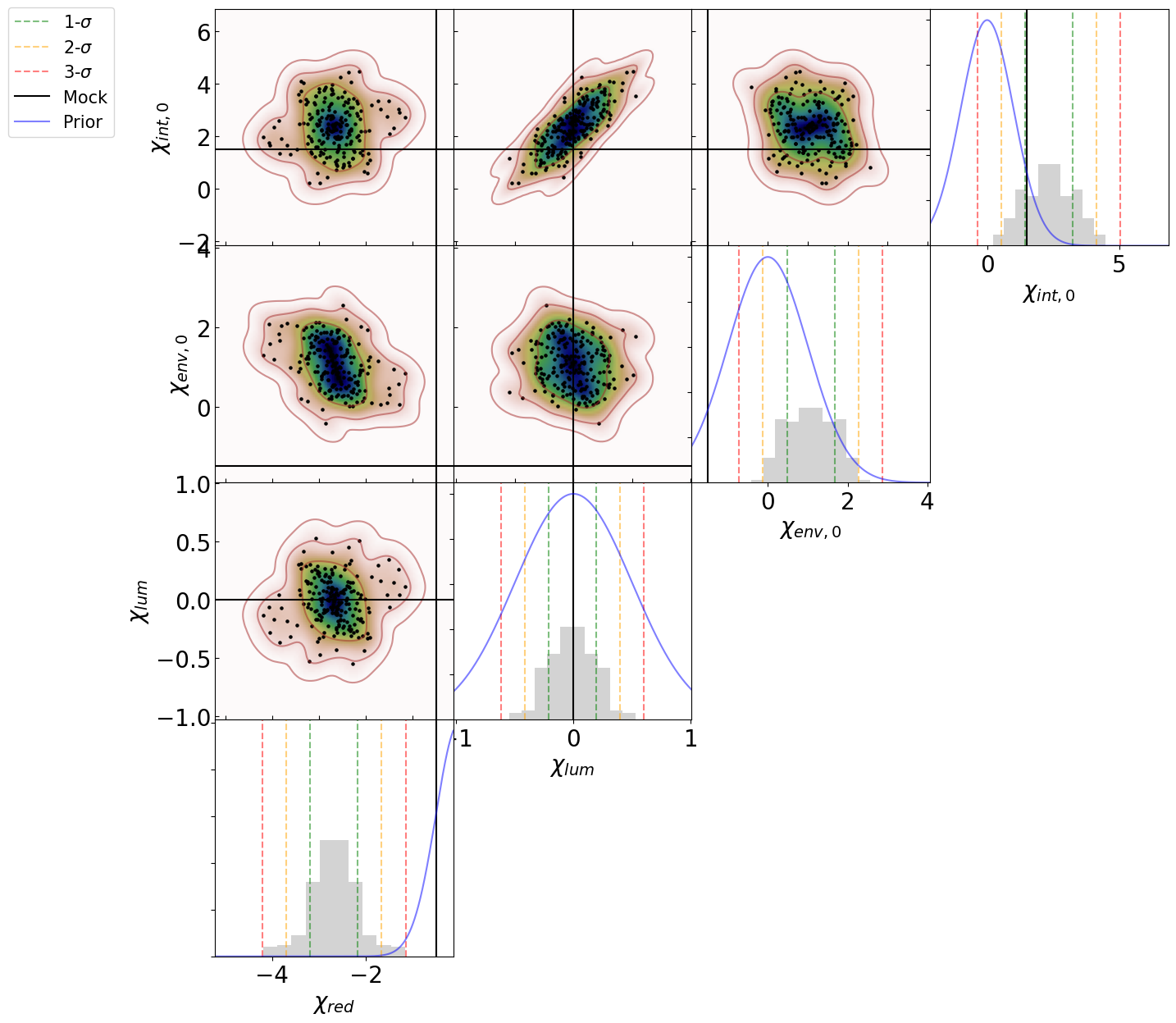}
    \caption{As in Fig.\,\ref{fig:skamid}, but for an extragalactic Faraday rotation standard deviation of 1.5\,rad/m$^2$ and using a catalogue including both mid- and low frequency observations.}

    \label{fig:skalow}
\end{figure}

Concerning the Galactic term, we produced synthetic rotation measure values by multiplying the dispersion measure image by \cite{Hutschenreuter2024} by a Gaussian random magnetic field with zero-mean and standard deviation equal to one. This magnetic field has been generated by using the following power spectrum $|B_{\ell}|^2$ as a function of wave number $\ell$,
\begin{equation}
|B_{\ell}|^2=\frac{P_0}{1.0 + \left( \frac{\ell}{\ell_0} \right) ^{-\gamma}},
\label{b_ps}
\end{equation}
where $P_0$ is the normalization of the power spectrum, $\ell_0$ is the characteristic wave number, and $\gamma$ is the slope of the power spectrum. The slope of the magnetic field power spectrum has been assumed to be $\gamma=-4$. The resulting image has been properly normalized in order to have units of rad/m$^2$ and is characterized by simultaneous fluctuations on small and large spatial scales. An example is shown in Fig.\,\ref{fig:galphi}. 
 We opted for this value of the slope of the magnetic field power spectrum because of the resemblance with the Galactic Faraday map obtained by \cite{Hutschenreuter2020} including data from free-free emission.

For a fraction of sources, we add an extragalactic term, randomly drawn for each line of sight from a zero-mean Gaussian with variance as given by Eq.\,\ref{egcontr}. The values of the $\Theta$ parameters have been fixed in order to produce overall extragalactic Faraday rotations in agreement with observations:
\begin{itemize}
\item $\approx$10\,rad/m$^2$ at mid-frequencies;
\item $\approx$1.5\,rad/m$^2$ at low frequencies.
\end{itemize}
The different magnitude of the extragalactic Faraday rotation effect simulated at mid- and low- frequencies, is due to the fact that the Faraday rotation effect at low frequencies is dominated by low-density weakly magnetized environments and has been shown to be characterized by a standard deviation approximately of 1.5\,rad/m$^2$ \citep{Carretti2022}. Mid-frequency observations carry information from denser and more magnetized environments, as galaxy clusters, with typical dispersion in Faraday rotation of about 10\,rad/m$^2$ \citep[e.g.,][]{Clarke2004}.

Due to computational limitations, we do not consider all the lines of sight that will be available at low- and mid- frequencies according to \S\,\ref{sourcecounts}. At mid-frequencies, we consider in total about 220,000 sources over the full sky (the most being located in the southern sky): around 160,000 of them have been used in order to constrain the Galactic contribution only, while about 50,000 to infer also the extragalactic Faraday rotation (about a few percent of the polarized sources expected to be available). Since the sensitivity in polarization expected for the survey assumed in this work is consistent with MeerKAT observations used by \cite{Loi2025}\footnote{\cite{Loi2025} observed a field of 6.35\,deg$^2$ in the 0-9-1.4\,GHz range and detected 508 polarized sources, which corresponds to a source density of 80 polarized sources per square degree.}, all the sources have been assumed to be characterized by a Faraday rotation uncertainty consistent with values given in that work, see top panel of Fig.\,\ref{fig:sigma_rm}.  

At low frequencies we use a similar setup. We consider a total of 210,000 sources over the full sky, about 193,000 of which have been used to constrain the Galactic contribution only and have a Faraday rotation uncertainty comparable with the data of \cite{Loi2025}. The remaining 17,000 (about 30\,percent of the number we expect will be available) have been exploited to infer also the extragalactic Faraday rotation and have been assumed to be characterized by a Faraday rotation uncertainty consistent with LOFAR RM data by \cite{OSullivan2023}, see top panel Fig.\,\ref{fig:sigma_rm}.

The sources used to assess the extragalactic Faraday rotation contribution are required to have auxiliary information (spectroscopic redshift and luminosity). Synthetic spectroscopic redshift and luminosity have been drawn from the observed distributions, see 
\cite{VanEck2023} and \cite{OSullivan2023}. 
In order to minimize the Galactic contamination, we assume that the Galactic latitude of these sources is higher than 45$^{\circ}$ in absolute value. 

\subsection{Results and discussion}
\label{results}
In Fig.\,\ref{fig:skamid}, we show the results of applying our algorithm to a RM catalogue corresponding to observations at mid-frequencies, assuming an overall extragalactic Faraday rotation of $\approx$10\,rad/m$^2$. In the top panels, we show 
 the mean and standard deviation of the Galactic Faraday sky posterior distribution, and in the bottom panels the posterior distribution of the extragalactic parameters. When compared with Fig.\,\ref{fig:galphi}, the reconstruction of the Galactic Faraday sky (top left panel in Fig.\,\ref{fig:skamid}) shows a
 high level of fidelity, with larger uncertainties in the northern sky where the sampling is poorer. The reconstructed Galactic Faraday sky looks smoother than the ground truth, likely because of a degeneracy between the Galactic contribution and the extragalactic and noise term at small-spatial scales. Concerning the extragalactic term, 
 the algorithm is capable of recovering the correct extragalactic parameters ($\chi_{\rm int,0}$, $\chi_{\rm env,0}$, $\chi_{\rm lum}$, and $\chi_{\rm red}$) in the synthetic data within one-sigma. We note a strong correlation between the parameters $\chi_{\rm int, 0}$ and $\chi_{\rm lum}$ and between the parameters $\chi_{\rm env, 0}$ and $\chi_{\rm red}$. Considering Eq.\,\ref{egcontr}, a degeneracy of these parameters can be expected: a given $\sigma_{\rm int}$ can be obtained for different combination of $\chi_{\rm int, 0}$ and $\chi_{\rm lum}$ values. Similarly, for $\sigma_{\rm env}$, $\chi_{\rm env, 0}$ and $\chi_{\rm red}$.

 The uncertainties in Faraday rotation adopted here reflect those derived by \cite{Loi2025} that do not include corrections for the ionospheric Faraday rotation. Resorting to long-term observations at the MeerKAT site, \cite{Hugo2024} find that residuals from ionospheric corrections can reach values up to $\sim$1\,rad/m$^2$. In the bottom panel of Fig.\,\ref{fig:sigma_rm} we show a comparison between Faraday rotation uncertainties derived by \cite{Loi2025} before and after including residuals from ionospheric rotation measure corrections as derived by \cite{Hugo2024}. Above 1\,rad/m$^2$, the two distributions do not significantly differ and their medians are comparable, being respectively $\approx$1.5\,rad/m$^2$ and $\approx$1.8\,rad/m$^2$, suggesting that uncertainties up to this level can be considered acceptable to our aims. Results by \cite{Taylor2024} 
show that the ionospheric stability is
better by a factor up to ten during night-time with respect to day-time observations, indicating that night-time observations must be preferred, especially when larger areas are observed with short pointings spread out over long times.

  In Fig.\,\ref{fig:skalow}, we show the extragalactic posterior by applying our algorithm to a rotation measure catalogue including both mid- and low frequencies data, and assuming an overall extragalactic Faraday rotation standard deviation of $\approx$1.5\,rad/m$^2$. The algorithm recovers $\chi_{\rm int, 0}$ and $\chi_{\rm lum}$ within one-sigma, and $\chi_{\rm env, 0}$ and $\chi_{\rm red}$ within four-sigma. While the extragalactic Faraday rotation is constrained only using low frequency data, to infer the Galactic sky both mid- and low frequency measurements are used. Differences, for a given source, in the Faraday rotation measured at mid- and low frequencies are expected to be primarily driven by the observational noise and/or due to depolarization effects that might render a polarized source unobservable at low frequencies or affect its Faraday depth spectrum \citep[see, for example, the 
recent work by][on RM Jitter]{Stil2025}. However, these differences should not have a significant impact on the inference of the Galactic Faraday rotation. Indeed, the Galactic term is constrained simultaneously using all the available rotation measure values, meaning that the Galactic Faraday rotation at a given position in the sky depends not only on the rotation measure observed in that direction, but also at close-by locations. 
We experimented different observing setups, e.g. catalogues including SKA-Mid and SKA-Low measurements only, as well as including NRAO VLA Sky Survey \citep[NVSS][]{Taylor2009} data points (not shown here). We find that catalogues from SKA-Mid and SKA-Low observations allow us to put better constraints on the Galactic Faraday rotation with respect to catalogues including also NVSS measurements, due to the larger (and sometimes unreliable) observational noise that affects NVSS rotation measures. 
Using only SKA-Low rotation measures, would be even better, due to the extremely small uncertainties on Faraday rotation, but the low density of polarized sources would not allow us a detailed view of the Galactic Faraday rotation, as that SKA-Mid catalogues are expected to deliver.

 Overall, to summarize, our algorithm shows better performances at mid-frequencies than at low frequencies, likely because of the larger magnitude of the extragalactic Faraday contribution we simulated, as described above. Larger samples will allow us to further reduce the uncertainty in the determination of extragalactic parameters, especially when sources at high absolute values of Galactic latitude are considered. 
 At low frequencies, a larger sample would require including sources closer to the Galactic plane. Increasing the number of sources for constraining extragalactic magnetization via the inclusion of lower latitude sources needs to be balanced against the contamination from Galactic small-scale structures at low latitudes.

\subsubsection{Spectroscopic versus photometric redshifts}
One of the major restrictions while studying Faraday rotation from extragalactic environments through rotation measure grids is the limited availability of spectroscopic redshifts. To date, spectroscopic redshifts have been identified for about 50\,percent of the polarized sources detected at low frequencies \citep[see][]{OSullivan2023} and for about 10\,percent of those currently available at mid-frequencies \citep[see][]{Hammond2012}. 
 We note that the incoming facilities are expected to provide spectroscopic redshifts for about $
10^8$ galaxies over 0 $<$ z $<$ 3 \citep[e.g.,][]{Mainieri2024}. However, despite the fact that their very small uncertainty allows us to locate sources along the line of sight with high accuracy, determining spectroscopic redshifts is extremely time-consuming, therefore the efforts of future optical surveys are mainly focused on delivering photometric redshifts \citep[e.g., ][]{Newman2022}.
Here, we want to shed light on the possibility to use photometric redshifts in order to constrain extragalactic Faraday rotation. 

\begin{figure}[h]
    \centering
        \includegraphics[width=1\columnwidth]{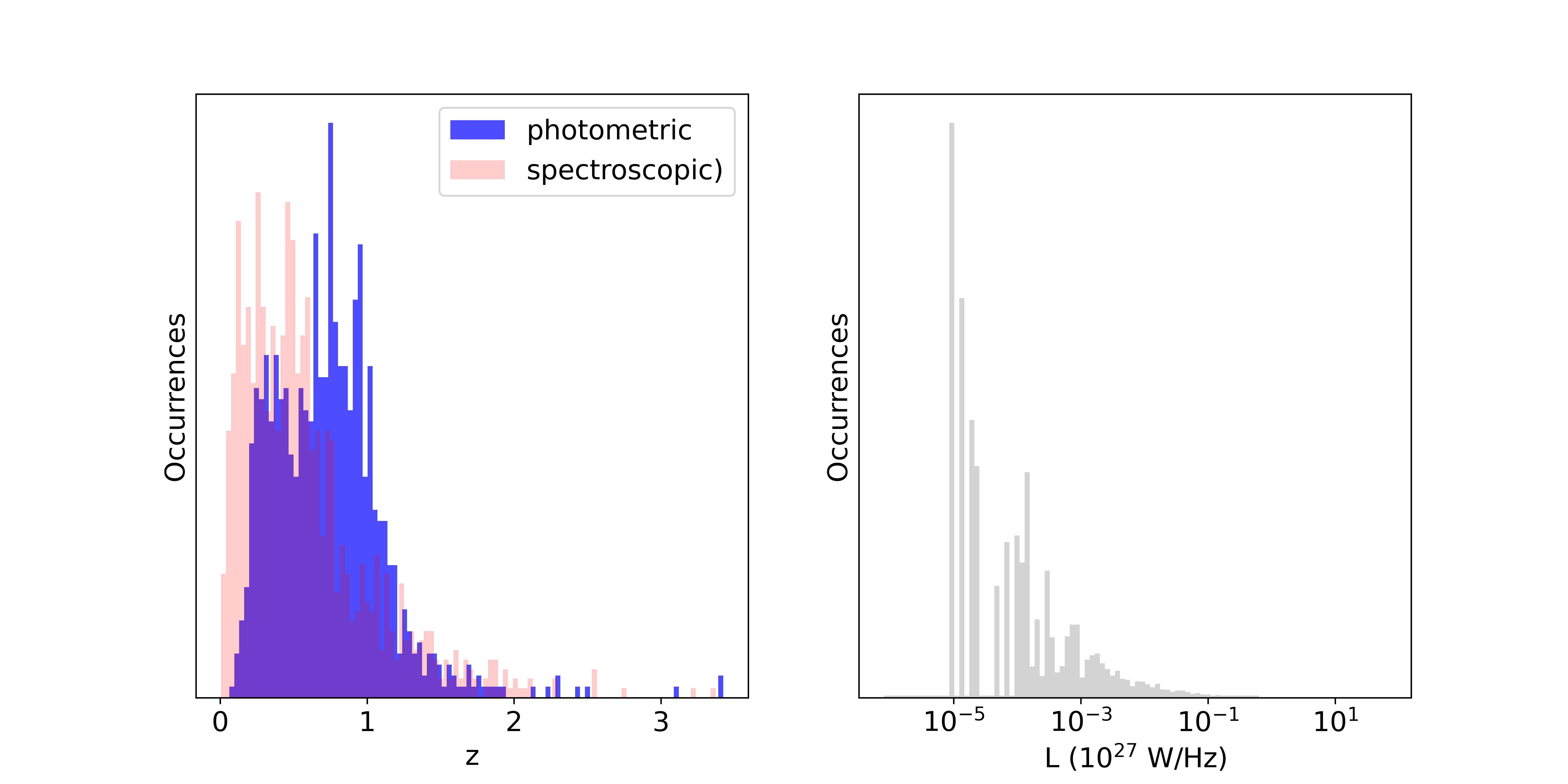}
       \caption{\emph{Left panel}: distribution of observed photometric (blue) and spectroscopic (red) redshifts. The purple colour shows the overlapping region. \emph{Right panel}: distribution of the luminosity of radio sources. Data have been taken from the LoTSS DR2 RM catalogue by \cite{OSullivan2023}, see text for more details. }
    \label{fig:zL}
\end{figure}

\begin{figure}[h]
    \centering
        	\includegraphics[width=\columnwidth]{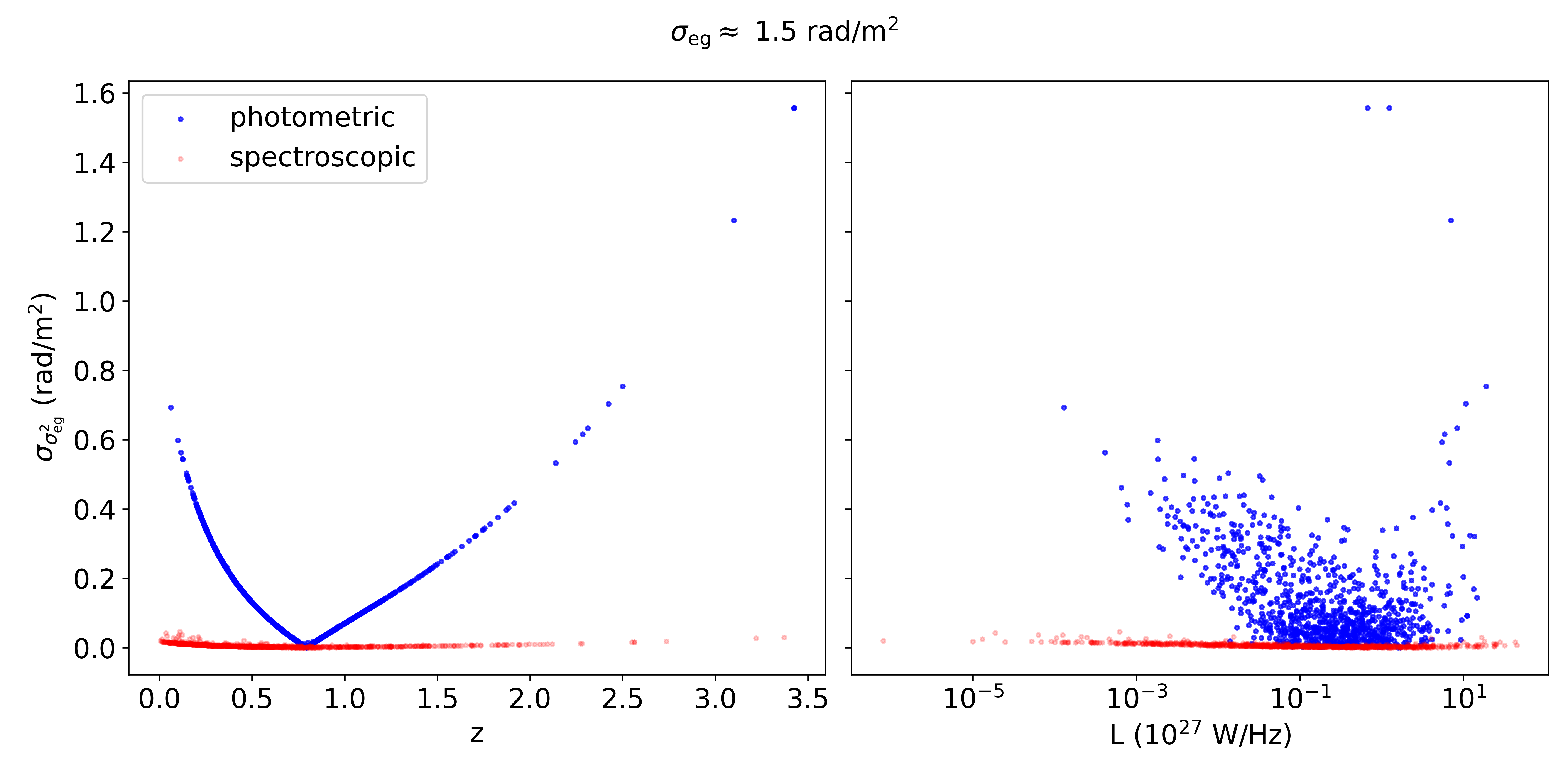}
                    	\includegraphics[width=\columnwidth]{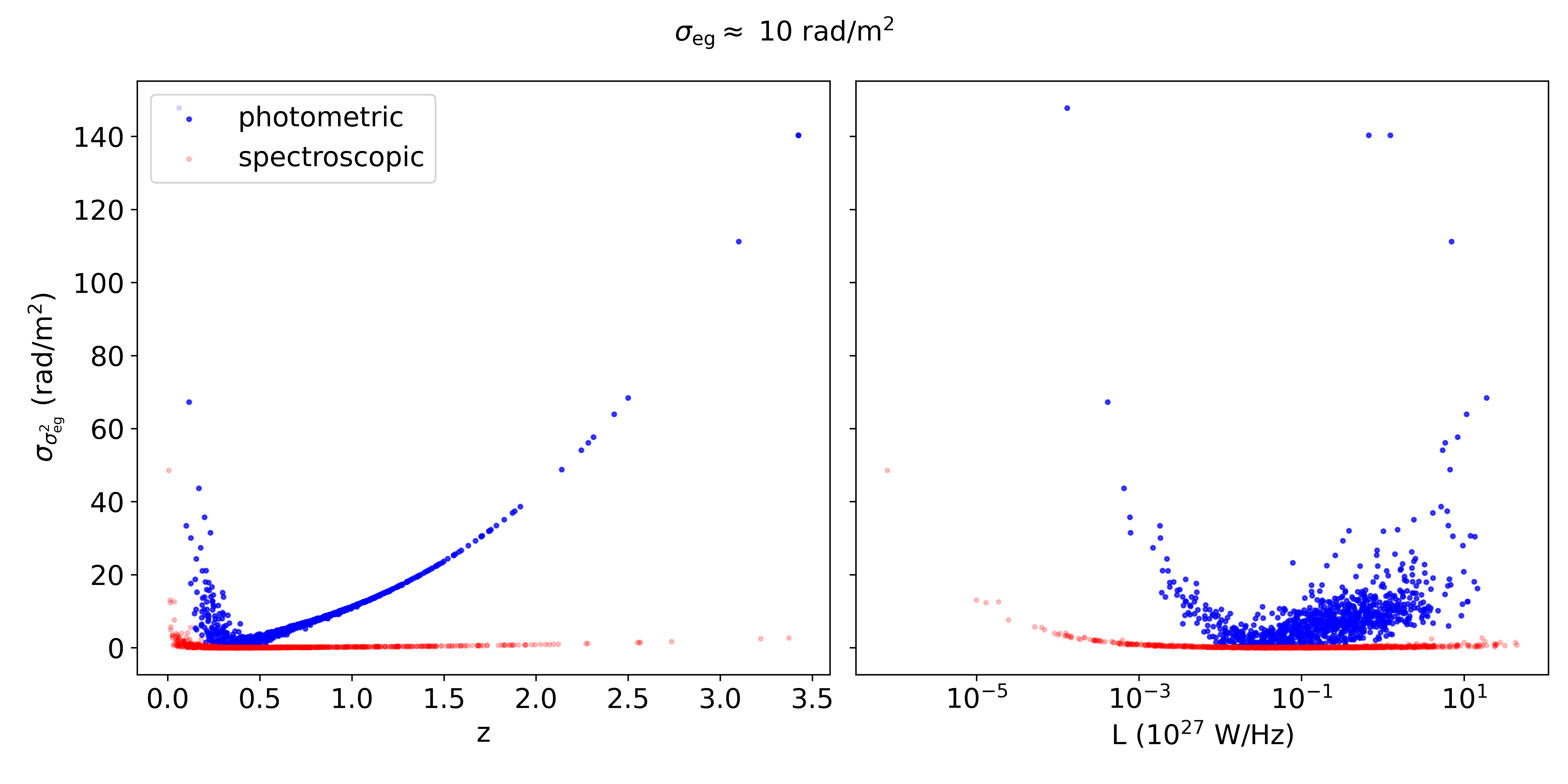}

       \caption{Uncertainty in Faraday rotation standard deviation as a function of redshift (\emph{left}) and luminosity (\emph{right}) of radio sources for spectroscopic (red) and photometric (blue) redshifts. On \emph{top panels}, the uncertainty in Faraday rotation has been estimated considering an overall extragalactic Faraday rotation standard deviation of about 1.5\,rad/m$^2$ while on \emph{bottom panels} of about 10\,rad/m$^2$. Redhsifts and luminosity values are the same shown in Fig.\,\ref{fig:zL}.}
    \label{fig:zaccuracy}
\end{figure}

 Labeling $\sigma_{\rm z}$ the uncertainty in redshift and $\sigma_{\rm S_{\nu}}$ the uncertainty in flux density $S_{\nu}$ at frequency $\nu$, the impact of these uncertainties on the detection of the extragalactic Faraday rotation variance given our model is
\begin{equation}
\begin{split}
\sigma_{\sigma_{\rm eg}^2}^2\approx \left|\frac{\partial{\sigma_{\rm eg}^2}}{\partial{z}} \right|^2\sigma_{\rm z}^2+\left|\frac{\partial{\sigma_{\rm eg}^2}}{\partial{S_{\nu}}} \right|^2\sigma_{\rm S_{\nu}}^2=\\
&
=\left [4(\chi_{\rm lum}-1)\left(\frac{L}{L_{\rm i}}\right)^{\chi_{\rm lum}}\frac{e^{\chi_{\rm int, 0}}}{(1+z)^5}+e^{\chi_{\rm env,0}} \frac{c}{H(z)}\frac{(1+z)^{4+\chi_{\rm red}}}{D_0}\right]^2\sigma_{\rm z}^2+\\
&
+\left[\frac{4\pi D_{\rm L}^2}{L_0}\left(\frac{L_{\rm i}}{L_0}\right)^{(\chi_{\rm lum}-1)}\frac{e^{\chi_{\rm int, 0}}}{(1+z)^4}\right]^2\sigma_{\rm S_{\nu}}^2,
\end{split}
\end{equation}
where $\sigma^2_{\rm eg}$ is given in Eq.\,\ref{egcontr}. The overall uncertainty 
depends on the redshift and on the flux density themself and on their uncertainties. The luminosity $L$ is related to the flux density by $L=4\pi (1+z)^{3+\alpha}D_{\rm A}^2 S_{\nu}$, where $D_{\rm A}$ is the angular-diameter distance, $\alpha$ is the spectral index\footnote{Following \cite{OSullivan2023} we assume $\alpha=0.7$ for all the sources, where the flux density $S_{\nu}$ at the frequency $\nu$ is defined as $S_{\nu}\propto\nu^{-\alpha}$.}, and the \emph{k}-correction has been taken into account.

For demonstration purposes, we consider the LoTSS DR2 RM catalogue by \cite{OSullivan2023}, where redshifts are available for about 2,000 sources, half of which are photometric and half spectroscopic. As described in \cite{Newman2022}, for photometric redshifts, we assume that
$\sigma_{\rm z}=0.05(1+z)$ represents a good lower limit for the redshift uncertainty with current surveys when different galaxy populations are considered. Concerning spectroscopic redshifts, we assume an uncertainty $\sigma_{\rm z}=0.001(1+z)$, in agreement with what expected for the Euclid survey \citep[see][and references therein]{Mauri2020}. 

In Fig.\,\ref{fig:zL}, we show the distributions of redshifts (spectroscopic and photometric) and of radio luminosities of the sources, considering the redshift and Stokes I\footnote{We assume that the polarized sources considered here are point-like. Therefore, Stokes I values have been assumed to correspond to flux densities.} measurements provided by \cite{OSullivan2023}.  In Fig.\,\ref{fig:zaccuracy},
we plot $\sigma_{\sigma_{\rm eg}^2}$ versus redshift and luminosity, both considering a set of $\Theta$ values corresponding to $\approx$\,1.5\,rad/m$^2$ (top) and to $\approx$\,10\,rad/m$^2$ (bottom). In the first case, the expected median uncertainty
is 0.005\,rad/m$^2$ in case of spectroscopic redshifts and 0.09\,rad/m$^2$  for photometric redshifts, while in the second case 0.12\,rad/m$^2$ for spectroscopic redshifts and 7.0\,rad/m$^2$ for photometric redshifts. In all cases, the redshift uncertainty dominates over that one due to the flux density by at least an order of magnitude.
To understand if photometric redshifts can be exploited, we should compare these numbers with the uncertainties in Faraday rotation measurements. Considering that the uncertainty in Faraday rotation estimates from low frequency observations is dominated by the ionospheric RM correction error and this amounts to  $\approx$\,0.06\,rad/m$^2$, spectroscopic redshifts are required when these catalogs are exploited, i.e. for the study of the Faraday rotation effect associated with the cosmic web. When considering observations at mid-frequencies, instead, a median uncertainty in Faraday rotation between $\approx$\,11\,rad/m$^2$ \citep[as in the case of shallow surveys as NVSS, see][]{Taylor2009} and 1.5-1.8\,rad/m$^2$ \citep[for deep observations see, e.g.,][ as discussed in \S\,\ref{results}, and for the polarization survey planned with SKA-Mid]{Loi2025} can be obtained. At mid-frequencies, in case of shallow surveys, photometric redshifts can be used without significantly affecting the characterization of the extragalactic Faraday rotation, whereas for deep observations and for the polarization survey planned with SKA-Mid, photometric uncertainty can not be considered negligible or comparable with respect to other involved uncertainties. In this case, spectroscopic redshifts should be used or, alternatively, the photometric redshift uncertainty should be properly taken into account.

\section{Conclusions}
\label{conclusions}
Shedding light on cosmic magnetism and it genesis requires a proper and detailed characterization of extragalactic magnetic fields. To make this possible, a 
reliable and accurate reconstruction of the Galactic Faraday sky is essential. We presented expectations for synthetic catalogues of Faraday rotation values for background point-like polarized extragalactic radio sources corresponding to SKA AA4 telescopes observations. To this end, we applied a Bayesian algorithm designed to simultaneously disentangle the Faraday rotation effect due to Galactic and extragalactic environments, properly taking into account the observing noise. We show that precise redshift information is critical in order to locate sources along the line of sight and track the path of their signal through extragalactic environments to the observer with negligible uncertainties. Spectroscopic redshifts are essential when dealing with the characterization of magnetization of the cosmic web, through low frequency observations, and of galaxy clusters with surveys planned with the SKA-Low and Mid telescopes. However, intracluster magnetic fields can be studied also by means of photometric redshifts when data from shallow NVSS-like radio surveys are exploited. 

\subsubsection*{Acknowledgements}
We thank the referee for the valuable comments and suggestions which helped improve the manuscript. 
V.\ V.\ acknowledges support from the Prize for Young Researchers "Gianni Tofani" second edition, promoted by INAF-Osservatorio Astrofisico di Arcetri (DD n. 84/2023).
S.\ H.\ acknowledges funding by the European Union (ERC, ISM-FLOW, 101055318). Views and opinions expressed are, however, those of the author(s) only and do not necessarily reflect those of the European Union or the European Research Council. Neither the European Union nor the granting authority can be held responsible for them. 
J.\ R.\ acknowledges financial support from the German Federal Ministry of Education and Research (BMBF) under grant 05A23WO1 (Verbundprojekt D-MeerKAT III).
S.\ P.\ O.\ acknowledges support from the Comunidad de Madrid Atracción de Talento program via grant 2022-T1/TIC-23797, and grant PID2023-146372OB-I00 funded by MICIU/AEI/10.13039/501100011033 and by ERDF, EU. This work was carried out thanks to the funding of the Regione Autonoma della Sardegna, ai sensi della Legge Regionale 7 agosto 2007, n.7 "Promozione della Ricerca Scientifica e dell'Innovazione Tecnologica in Sardegna".

\bibliographystyle{abbrvnat-maxbibnames4}
\bibliography{chapter} 

@ARTICLE{Burn1966,
       author = {{Burn}, B.~J.},
        title = "{On the depolarization of discrete radio sources by Faraday dispersion}",
      journal = {MNRAS},
         year = 1966,
        month = jan,
       volume = {133},
        pages = {67},
          doi = {10.1093/mnras/133.1.67},
       adsurl = {https://ui.adsabs.harvard.edu/abs/1966MNRAS.133...67B}
}

@ARTICLE{Carretti2022,
       author = {{Carretti}, Ettore and {Vacca}, V. and {O'Sullivan}, S.~P. and {Heald}, G.~H. and {Horellou}, C. and {R{\"o}ttgering}, H.~J.~A. and {Scaife}, A.~M.~M. and {Shimwell}, T.~W. and {Shulevski}, A. and {Stuardi}, C. and {Vernstrom}, T.},
        title = "{Magnetic field strength in cosmic web filaments}",
      journal = {MNRAS},
         year = 2022,
        month = may,
       volume = {512},
       number = {1},
        pages = {945-959},
          doi = {10.1093/mnras/stac384},
archivePrefix = {arXiv},
       eprint = {2202.04607},
 primaryClass = {astro-ph.CO},
       adsurl = {https://ui.adsabs.harvard.edu/abs/2022MNRAS.512..945C}
}

@ARTICLE{Carretti2023,
       author = {{Carretti}, E. and {O'Sullivan}, S.~P. and {Vacca}, V. and {Vazza}, F. and {Gheller}, C. and {Vernstrom}, T. and {Bonafede}, A.},
        title = "{Magnetic field evolution in cosmic filaments with LOFAR data}",
      journal = {MNRAS},
         year = 2023,
        month = jan,
       volume = {518},
       number = {2},
        pages = {2273-2286},
          doi = {10.1093/mnras/stac2966},
archivePrefix = {arXiv},
       eprint = {2210.06220},
 primaryClass = {astro-ph.CO},
       adsurl = {https://ui.adsabs.harvard.edu/abs/2023MNRAS.518.2273C}
}

@ARTICLE{Carretti2025,
       author = {{Carretti}, E. and {Vazza}, F. and {O'Sullivan}, S.~P. and {Vacca}, V. and {Bonafede}, A. and {Heald}, G. and {Horellou}, C. and {Mtchedlidze}, S. and {Vernstrom}, T.},
        title = "{The nature of LOFAR rotation measures and new constraints on magnetic fields in cosmic filaments and on magnetogenesis scenarios}",
      journal = {A\&A},
         year = 2025,
        month = jan,
       volume = {693},
          eid = {A208},
        pages = {A208},
          doi = {10.1051/0004-6361/202451333},
archivePrefix = {arXiv},
       eprint = {2411.13499},
 primaryClass = {astro-ph.CO},
       adsurl = {https://ui.adsabs.harvard.edu/abs/2025A&A...693A.208C}
}

@ARTICLE{Clarke2004,
       author = {{Clarke}, Tracy E.},
        title = "{Faraday Rotation Observations of Magnetic Fields in Galaxy Clusters}",
      journal = {Journal of Korean Astronomical Society},
         year = 2004,
        month = dec,
       volume = {37},
       number = {5},
        pages = {337-342},
          doi = {10.5303/JKAS.2004.37.5.337},
archivePrefix = {arXiv},
       eprint = {astro-ph/0412268},
 primaryClass = {astro-ph},
       adsurl = {https://ui.adsabs.harvard.edu/abs/2004JKAS...37..337C}
}

@ARTICLE{Ensslin2009,
       author = {{En{\ss}lin}, Torsten A. and {Frommert}, Mona and {Kitaura}, Francisco S.},
        title = "{Information field theory for cosmological perturbation reconstruction and nonlinear signal analysis}",
      journal = {Physical Review D},
         year = 2009,
        month = nov,
       volume = {80},
       number = {10},
          eid = {105005},
        pages = {105005},
          doi = {10.1103/PhysRevD.80.105005},
archivePrefix = {arXiv},
       eprint = {0806.3474},
 primaryClass = {astro-ph},
       adsurl = {https://ui.adsabs.harvard.edu/abs/2009PhRvD..80j5005E}
}

@ARTICLE{Frank2021,
       author = {{Frank}, Philipp and {Leike}, Reimar and {En{\ss}lin}, Torsten A.},
        title = "{Geometric Variational Inference}",
      journal = {Entropy},
         year = 2021,
        month = jul,
       volume = {23},
       number = {7},
          eid = {853},
        pages = {853},
          doi = {10.3390/e23070853},
archivePrefix = {arXiv},
       eprint = {2105.10470},
 primaryClass = {stat.ME},
       adsurl = {https://ui.adsabs.harvard.edu/abs/2021Entrp..23..853F}
}

@ARTICLE{Govoni2004,
       author = {{Govoni}, Federica and {Feretti}, Luigina},
        title = "{Magnetic Fields in Clusters of Galaxies}",
      journal = {International Journal of Modern Physics D},
         year = 2004,
        month = jan,
       volume = {13},
       number = {8},
        pages = {1549-1594},
          doi = {10.1142/S0218271804005080},
archivePrefix = {arXiv},
       eprint = {astro-ph/0410182},
 primaryClass = {astro-ph},
       adsurl = {https://ui.adsabs.harvard.edu/abs/2004IJMPD..13.1549G}
}

@ARTICLE{Govoni2019,
       author = {{Govoni}, F. and {Orr{\`u}}, E. and {Bonafede}, A. and {Iacobelli}, M. and {Paladino}, R. and {Vazza}, F. and {Murgia}, M. and {Vacca}, V. and {Giovannini}, G. and {Feretti}, L. and {Loi}, F. and {Bernardi}, G. and {Ferrari}, C. and {Pizzo}, R.~F. and {Gheller}, C. and {Manti}, S. and {Br{\"u}ggen}, M. and {Brunetti}, G. and {Cassano}, R. and {de Gasperin}, F. and {En{\ss}lin}, T.~A. and {Hoeft}, M. and {Horellou}, C. and {Junklewitz}, H. and {R{\"o}ttgering}, H.~J.~A. and {Scaife}, A.~M.~M. and {Shimwell}, T.~W. and {van Weeren}, R.~J. and {Wise}, M.},
        title = "{A radio ridge connecting two galaxy clusters in a filament of the cosmic web}",
      journal = {Science},
         year = 2019,
        month = jun,
       volume = {364},
       number = {6444},
        pages = {981-984},
          doi = {10.1126/science.aat7500},
archivePrefix = {arXiv},
       eprint = {1906.07584},
 primaryClass = {astro-ph.GA},
       adsurl = {https://ui.adsabs.harvard.edu/abs/2019Sci...364..981G}
}

@ARTICLE{Hammond2012,
       author = {{Hammond}, Alison M. and {Robishaw}, Timothy and {Gaensler}, B.~M.},
        title = "{A New Catalog of Faraday Rotation Measures and Redshifts for Extragalactic Radio Sources}",
      journal = {arXiv e-prints},
         year = 2012,
        month = sep,
          eid = {arXiv:1209.1438},
        pages = {arXiv:1209.1438},
          doi = {10.48550/arXiv.1209.1438},
archivePrefix = {arXiv},
       eprint = {1209.1438},
 primaryClass = {astro-ph.CO},
       adsurl = {https://ui.adsabs.harvard.edu/abs/2012arXiv1209.1438H}
}

@ARTICLE{Heald2020,
       author = {{Heald}, George and {Mao}, Sui Ann and {Vacca}, Valentina and {Akahori}, Takuya and {Damas-Segovia}, Ancor and {Gaensler}, B.~M. and {Hoeft}, Matthias and {Agudo}, Ivan and {Basu}, Aritra and {Beck}, Rainer and {Birkinshaw}, Mark and {Bonafede}, Annalisa and {Bourke}, Tyler L. and {Bracco}, Andrea and {Carretti}, Ettore and {Feretti}, Luigina and {Girart}, J.~M. and {Govoni}, Federica and {Green}, James A. and {Han}, JinLin and {Haverkorn}, Marijke and {Horellou}, Cathy and {Johnston-Hollitt}, Melanie and {Kothes}, Roland and {Landecker}, Tom and {Nikiel-Wroczy{\'n}ski}, B{\l}a{\.z}ej and {O'Sullivan}, Shane P. and {Padovani}, Marco and {Poidevin}, Fr{\'e}d{\'e}rick and {Pratley}, Luke and {Regis}, Marco and {Riseley}, Christopher John and {Robishaw}, Tim and {Rudnick}, Lawrence and {Sobey}, Charlotte and {Stil}, Jeroen M. and {Sun}, Xiaohui and {Sur}, Sharanya and {Taylor}, A. Russ and {Thomson}, Alec and {Van Eck}, Cameron L. and {Vazza}, Franco and {West}, Jennifer L. and {the SKA Magnetism Science Working Group}},
        title = "{Magnetism Science with the Square Kilometre Array}",
      journal = {Galaxies},
         year = 2020,
        month = jul,
       volume = {8},
       number = {3},
          eid = {53},
        pages = {53},
          doi = {10.3390/galaxies8030053},
archivePrefix = {arXiv},
       eprint = {2006.03172},
 primaryClass = {astro-ph.GA},
       adsurl = {https://ui.adsabs.harvard.edu/abs/2020Galax...8...53H}
}

@ARTICLE{Hugo2024,
       author = {{Hugo}, B. and {Perley}, R.},
        title = "{Absolute linear polarization angle calibration using planetary bodies for
MeerKAT and JVLA at cm wavelengths}",
      journal = {},
         year = 2024,
        month = jan,
       volume = {},
          eid = {SSA-0004E-001},
        pages = {},
          doi = {},
archivePrefix = {},
       eprint = {},
 primaryClass = {},
       adsurl = {https://archive-gw-1.kat.ac.za/public/repository/10.48479/bqk7-aw53/data/Absolute_linear_polarization_angle_calibration_using_planetary_bodies_for_MeerKAT_and_JVLA-REVB.pdf}
}

@ARTICLE{Hutschenreuter2020,
       author = {{Hutschenreuter}, Sebastian and {En{\ss}lin}, Torsten A.},
        title = "{The Galactic Faraday depth sky revisited}",
      journal = {A\&A},
         year = 2020,
        month = jan,
       volume = {633},
          eid = {A150},
        pages = {A150},
          doi = {10.1051/0004-6361/201935479},
archivePrefix = {arXiv},
       eprint = {1903.06735},
 primaryClass = {astro-ph.GA},
       adsurl = {https://ui.adsabs.harvard.edu/abs/2020A&A...633A.150H}
}

@ARTICLE{Hutschenreuter2022,
       author = {{Hutschenreuter}, S. and {Anderson}, C.~S. and {Betti}, S. and {Bower}, G.~C. and {Brown}, J. -A. and {Br{\"u}ggen}, M. and {Carretti}, E. and {Clarke}, T. and {Clegg}, A. and {Costa}, A. and {Croft}, S. and {Van Eck}, C. and {Gaensler}, B.~M. and {de Gasperin}, F. and {Haverkorn}, M. and {Heald}, G. and {Hull}, C.~L.~H. and {Inoue}, M. and {Johnston-Hollitt}, M. and {Kaczmarek}, J. and {Law}, C. and {Ma}, Y.~K. and {MacMahon}, D. and {Mao}, S.~A. and {Riseley}, C. and {Roy}, S. and {Shanahan}, R. and {Shimwell}, T. and {Stil}, J. and {Sobey}, C. and {O'Sullivan}, S.~P. and {Tasse}, C. and {Vacca}, V. and {Vernstrom}, T. and {Williams}, P.~K.~G. and {Wright}, M. and {En{\ss}lin}, T.~A.},
        title = "{The Galactic Faraday rotation sky 2020}",
      journal = {A\&A},
         year = 2022,
        month = jan,
       volume = {657},
          eid = {A43},
        pages = {A43},
          doi = {10.1051/0004-6361/202140486},
archivePrefix = {arXiv},
       eprint = {2102.01709},
 primaryClass = {astro-ph.GA},
       adsurl = {https://ui.adsabs.harvard.edu/abs/2022A&A...657A..43H}
}

@ARTICLE{Hutschenreuter2024,
       author = {{Hutschenreuter}, Sebastian and {Haverkorn}, Marijke and {Frank}, Philipp and {Raycheva}, Nergis C. and {En{\ss}lin}, Torsten A.},
        title = "{Disentangling the Faraday rotation sky}",
      journal = {A\&A},
         year = 2024,
        month = oct,
       volume = {690},
          eid = {A314},
        pages = {A314},
          doi = {10.1051/0004-6361/202346740},
archivePrefix = {arXiv},
       eprint = {2304.12350},
 primaryClass = {astro-ph.GA},
       adsurl = {https://ui.adsabs.harvard.edu/abs/2024A&A...690A.314H}
}

@ARTICLE{Kronberg2008,
       author = {{Kronberg}, P.~P. and {Bernet}, M.~L. and {Miniati}, F. and {Lilly}, S.~J. and {Short}, M.~B. and {Higdon}, D.~M.},
        title = "{A Global Probe of Cosmic Magnetic Fields to High Redshifts}",
      journal = {ApJ},
         year = 2008,
        month = mar,
       volume = {676},
       number = {1},
        pages = {70-79},
          doi = {10.1086/527281},
archivePrefix = {arXiv},
       eprint = {0712.0435},
 primaryClass = {astro-ph},
       adsurl = {https://ui.adsabs.harvard.edu/abs/2008ApJ...676...70K}
}

@ARTICLE{Lin2024,
       author = {{Lin}, Jinyang and {Zhu}, Zhenghao and {Ma}, Renyi and {Bonaldi}, Anna and {Shan}, Huanyuan},
        title = "{A new model for the extragalactic radio sky at low frequency calibrated using the LOFAR Two-metre Survey}",
      journal = {RAS Techniques and Instruments},
         year = 2024,
        month = jan,
       volume = {3},
       number = {1},
        pages = {737-747},
          doi = {10.1093/rasti/rzae051},
archivePrefix = {arXiv},
       eprint = {2411.03931},
 primaryClass = {astro-ph.GA},
       adsurl = {https://ui.adsabs.harvard.edu/abs/2024RASTI...3..737L}
}

@ARTICLE{Loi2025,
       author = {{Loi}, F. and {Serra}, P. and {Murgia}, M. and {Govoni}, F. and {Vacca}, V. and {Maccagni}, F. and {Kleiner}, D. and {Kamphuis}, P.},
        title = "{The MeerKAT Fornax Survey: IV. A close look at the cluster physics through the densest rotation measure grid}",
      journal = {A\&A},
         year = 2025,
        month = feb,
       volume = {694},
          eid = {A125},
        pages = {A125},
          doi = {10.1051/0004-6361/202451711},
archivePrefix = {arXiv},
       eprint = {2501.05519},
 primaryClass = {astro-ph.CO},
       adsurl = {https://ui.adsabs.harvard.edu/abs/2025A&A...694A.125L}
}

@ARTICLE{Mainieri2024,
       author = {{Mainieri}, Vincenzo and {Anderson}, Richard I. and {Brinchmann}, Jarle and {Cimatti}, Andrea and {Ellis}, Richard S. and {Hill}, Vanessa and {Kneib}, Jean-Paul and {McLeod}, Anna F. and {Opitom}, Cyrielle and {Roth}, Martin M. and {Sanchez-Saez}, Paula and {Smiljanic}, Rodolfo and {Tolstoy}, Eline and {Bacon}, Roland and {Randich}, Sofia and {Adamo}, Angela and {Annibali}, Francesca and {Arevalo}, Patricia and {Audard}, Marc and {Barsanti}, Stefania and {Battaglia}, Giuseppina and {Bayo Aran}, Amelia M. and {Belfiore}, Francesco and {Bellazzini}, Michele and {Bellini}, Emilio and {Beltran}, Maria Teresa and {Berni}, Leda and {Bianchi}, Simone and {Biazzo}, Katia and {Bisero}, Sofia and {Bisogni}, Susanna and {Bland-Hawthorn}, Joss and {Blondin}, Stephane and {Bodensteiner}, Julia and {Boffin}, Henri M.~J. and {Bonito}, Rosaria and {Bono}, Giuseppe and {Bouche}, Nicolas F. and {Bowman}, Dominic and {Braga}, Vittorio F. and {Bragaglia}, Angela and {Branchesi}, Marica and {Brucalassi}, Anna and {Bryant}, Julia J. and {Bryson}, Ian and {Busa}, Innocenza and {Camera}, Stefano and {Carbone}, Carmelita and {Casali}, Giada and {Casali}, Mark and {Casasola}, Viviana and {Castro}, Norberto and {Catelan}, Marcio and {Cavallo}, Lorenzo and {Chiappini}, Cristina and {Cioni}, Maria-Rosa and {Colless}, Matthew and {Colzi}, Laura and {Contarini}, Sofia and {Couch}, Warrick and {D'Ammando}, Filippo and {d'Assignies D.}, William and {D'Orazi}, Valentina and {da Silva}, Ronaldo and {Dainotti}, Maria Giovanna and {Damiani}, Francesco and {Danielski}, Camilla and {De Cia}, Annalisa and {de Jong}, Roelof S. and {Dhawan}, Suhail and {Dierickx}, Philippe and {Driver}, Simon P. and {Dupletsa}, Ulyana and {Escoffier}, Stephanie and {Escorza}, Ana and {Fabrizio}, Michele and {Fiorentino}, Giuliana and {Fontana}, Adriano and {Fontani}, Francesco and {Forero Sanchez}, Daniel and {Franois}, Patrick and {Galindo-Guil}, Francisco Jose and {Gallazzi}, Anna Rita and {Galli}, Daniele and {Garcia}, Miriam and {Garcia-Rojas}, Jorge and {Garilli}, Bianca and {Grand}, Robert and {Guarcello}, Mario Giuseppe and {Hazra}, Nandini and {Helmi}, Amina and {Herrero}, Artemio and {Iglesias}, Daniela and {Ilic}, Dragana and {Irsic}, Vid and {Ivanov}, Valentin D. and {Izzo}, Luca and {Jablonka}, Pascale and {Joachimi}, Benjamin and {Kakkad}, Darshan and {Kamann}, Sebastian and {Koposov}, Sergey and {Kordopatis}, Georges and {Kovacevic}, Andjelka B. and {Kraljic}, Katarina and {Kuncarayakti}, Hanindyo and {Kwon}, Yuna and {La Forgia}, Fiorangela and {Lahav}, Ofer and {Laigle}, Clotilde and {Lazzarin}, Monica and {Leaman}, Ryan and {Leclercq}, Floriane and {Lee}, Khee-Gan and {Lee}, David and {Lehnert}, Matt D. and {Lira}, Paulina and {Loffredo}, Eleonora and {Lucatello}, Sara and {Magrini}, Laura and {Maguire}, Kate and {Mahler}, Guillaume and {Zahra Majidi}, Fatemeh and {Malavasi}, Nicola and {Mannucci}, Filippo and {Marconi}, Marcella and {Martin}, Nicolas and {Marulli}, Federico and {Massari}, Davide and {Matsuno}, Tadafumi and {Mattheee}, Jorryt and {McGee}, Sean and {Merc}, Jaroslav and {Merle}, Thibault and {Miglio}, Andrea and {Migliorini}, Alessandra and {Minchev}, Ivan and {Minniti}, Dante and {Miret-Roig}, Nuria and {Monreal Ibero}, Ana and {Montano}, Federico and {Montet}, Ben T. and {Moresco}, Michele and {Moretti}, Chiara and {Moscardini}, Lauro and {Moya}, Andres and {Mueller}, Oliver and {Nanayakkara}, Themiya and {Nicholl}, Matt and {Nordlander}, Thomas and {Onori}, Francesca and {Padovani}, Marco and {Pala}, Anna Francesca and {Panda}, Swayamtrupta and {Pandey-Pommier}, Mamta and {Pasquini}, Luca and {Pawlak}, Michal and {Pessi}, Priscila J. and {Pisani}, Alice and {Popovic}, Lukav C. and {Prisinzano}, Loredana and {Raddi}, Roberto and {Rainer}, Monica and {Rebassa-Mansergas}, Alberto and {Richard}, Johan and {Rigault}, Mickael and {Rocher}, Antoine and {Romano}, Donatella and {Rosati}, Piero and {Sacco}, Germano and {Sanchez-Janssen}, Ruben and {Sander}, Andreas A.~C. and {Sanders}, Jason L. and {Sargent}, Mark and {Sarpa}, Elena and {Schimd}, Carlo and {Schipani}, Pietro and {Sefusatti}, Emiliano and {Smith}, Graham P. and {Spina}, Lorenzo and {Steinmetz}, Matthias and {Tacchella}, Sandro and {Tautvaisiene}, Grazina and {Theissen}, Christopher and {Thomas}, Guillaume and {Ting}, Yuan-Sen and {Travouillon}, Tony and {Tresse}, Laurence and {Trivedi}, Oem and {Tsantaki}, Maria and {Tsedrik}, Maria and {Urrutia}, Tanya and {Valenti}, Elena and {Van der Swaelmen}, Mathieu and {Van Eck}, Sophie and {Verdiani}, Francesco and {Verdier}, Aurelien and {Vergani}, Susanna Diana and {Verhamme}, Anne and {Vernet}, Joel},
        title = "{The Wide-field Spectroscopic Telescope (WST) Science White Paper}",
      journal = {arXiv e-prints},
         year = 2024,
        month = mar,
          eid = {arXiv:2403.05398},
        pages = {arXiv:2403.05398},
          doi = {10.48550/arXiv.2403.05398},
archivePrefix = {arXiv},
       eprint = {2403.05398},
 primaryClass = {astro-ph.IM},
       adsurl = {https://ui.adsabs.harvard.edu/abs/2024arXiv240305398M}
}

@INPROCEEDINGS{Mauri2020,
       author = {{Mauri}, N. and {Dusini}, S. and {Fornari}, F. and {Di Ferdinando}, D. and {Giacomini}, F. and {Laudisio}, F. and {Patrizii}, L. and {Sirignano}, C. and {Sirri}, G. and {Stanco}, L. and {Tenti}, M. and {Valenziano}, L. and {Consortium}, Euclid},
        title = "{The Euclid Near Infrared Spectro-Photometer (NISP) instrument and science}",
    booktitle = {Journal of Physics Conference Series},
         year = 2020,
       series = {Journal of Physics Conference Series},
       volume = {1342},
        month = jan,
    publisher = {IOP},
          eid = {012122},
        pages = {012122},
          doi = {10.1088/1742-6596/1342/1/012122},
       adsurl = {https://ui.adsabs.harvard.edu/abs/2020JPhCS1342a2122M}
}

@ARTICLE{Newman2022,
       author = {{Newman}, Jeffrey A. and {Gruen}, Daniel},
        title = "{Photometric Redshifts for Next-Generation Surveys}",
      journal = {ARA\&A},
         year = 2022,
        month = aug,
       volume = {60},
        pages = {363-414},
          doi = {10.1146/annurev-astro-032122-014611},
archivePrefix = {arXiv},
       eprint = {2206.13633},
 primaryClass = {astro-ph.CO},
       adsurl = {https://ui.adsabs.harvard.edu/abs/2022ARA&A..60..363N}
}

@ARTICLE{Oppermann2015,
       author = {{Oppermann}, N. and {Junklewitz}, H. and {Greiner}, M. and {En{\ss}lin}, T.~A. and {Akahori}, T. and {Carretti}, E. and {Gaensler}, B.~M. and {Goobar}, A. and {Harvey-Smith}, L. and {Johnston-Hollitt}, M. and {Pratley}, L. and {Schnitzeler}, D.~H.~F.~M. and {Stil}, J.~M. and {Vacca}, V.},
        title = "{Estimating extragalactic Faraday rotation}",
      journal = {A\&A},
         year = 2015,
        month = mar,
       volume = {575},
          eid = {A118},
        pages = {A118},
          doi = {10.1051/0004-6361/201423995},
archivePrefix = {arXiv},
       eprint = {1404.3701},
 primaryClass = {astro-ph.IM},
       adsurl = {https://ui.adsabs.harvard.edu/abs/2015A&A...575A.118O}
}

@ARTICLE{OSullivan2023,
       author = {{O'Sullivan}, S.~P. and {Shimwell}, T.~W. and {Hardcastle}, M.~J. and {Tasse}, C. and {Heald}, G. and {Carretti}, E. and {Br{\"u}ggen}, M. and {Vacca}, V. and {Sobey}, C. and {Van Eck}, C.~L. and {Horellou}, C. and {Beck}, R. and {Bilicki}, M. and {Bourke}, S. and {Botteon}, A. and {Croston}, J.~H. and {Drabent}, A. and {Duncan}, K. and {Heesen}, V. and {Ideguchi}, S. and {Kirwan}, M. and {Lawlor}, L. and {Mingo}, B. and {Nikiel-Wroczy{\'n}ski}, B. and {Piotrowska}, J. and {Scaife}, A.~M.~M. and {van Weeren}, R.~J.},
        title = "{The Faraday Rotation Measure Grid of the LOFAR Two-metre Sky Survey: Data Release 2}",
      journal = {MNRAS},
         year = 2023,
        month = mar,
       volume = {519},
       number = {4},
        pages = {5723-5742},
          doi = {10.1093/mnras/stac3820},
archivePrefix = {arXiv},
       eprint = {2301.07697},
 primaryClass = {astro-ph.CO},
       adsurl = {https://ui.adsabs.harvard.edu/abs/2023MNRAS.519.5723O}
}

@incollection{OSullivan2026, 
    author = {Shane P. O'Sullivan and author2 and author3 and author4 and author5},
    title = {},
    year = {2026},
    publisher = {},
    note = {arXiv search: Report number AASKAII/OSullivan01},
    booktitle = {Advancing Astrophysics with the SKA -- II (AASKAII)}}

@ARTICLE{Piras2024,
       author = {{Piras}, S. and {Horellou}, C. and {Conway}, J.~E. and {Thomasson}, M. and {del Palacio}, S. and {Shimwell}, T.~W. and {O'Sullivan}, S.~P. and {Carretti}, E. and {{\v{S}}nidari{\'c}}, I. and {Jeli{\'c}}, V. and {Adebahr}, B. and {Berger}, A. and {Best}, P.~N. and {Br{\"u}ggen}, M. and {Herrera Ruiz}, N. and {Paladino}, R. and {Prandoni}, I. and {Sabater}, J. and {Vacca}, V.},
        title = "{LOFAR Deep Fields: Probing the sub-mJy regime of polarized extragalactic sources in ELAIS-N1. I. The catalog}",
      journal = {A\&A},
         year = 2024,
        month = jul,
       volume = {687},
          eid = {A267},
        pages = {A267},
          doi = {10.1051/0004-6361/202349085},
archivePrefix = {arXiv},
       eprint = {2406.08346},
 primaryClass = {astro-ph.CO},
       adsurl = {https://ui.adsabs.harvard.edu/abs/2024A&A...687A.267P}
}

@ARTICLE{Rudnick2014,
       author = {{Rudnick}, L. and {Owen}, F.~N.},
        title = "{The Distribution of Polarized Radio Sources >15 {\ensuremath{\mu}}Jy in GOODS-N}",
      journal = {ApJ},
         year = 2014,
        month = apr,
       volume = {785},
       number = {1},
          eid = {45},
        pages = {45},
          doi = {10.1088/0004-637X/785/1/45},
archivePrefix = {arXiv},
       eprint = {1402.3637},
 primaryClass = {astro-ph.GA},
       adsurl = {https://ui.adsabs.harvard.edu/abs/2014ApJ...785...45R}
}

@ARTICLE{Schnitzeler2010,
       author = {{Schnitzeler}, D.~H.~F.~M.},
        title = "{The latitude dependence of the rotation measures of NVSS sources}",
      journal = {MNRAS},
         year = 2010,
        month = nov,
       volume = {409},
       number = {1},
        pages = {L99-L103},
          doi = {10.1111/j.1745-3933.2010.00957.x},
archivePrefix = {arXiv},
       eprint = {1011.0737},
 primaryClass = {astro-ph.GA},
       adsurl = {https://ui.adsabs.harvard.edu/abs/2010MNRAS.409L..99S}
}

@ARTICLE{Stil2014,
       author = {{Stil}, J.~M. and {Keller}, B.~W. and {George}, S.~J. and {Taylor}, A.~R.},
        title = "{Degree of Polarization and Source Counts of Faint Radio Sources from Stacking Polarized Intensity}",
      journal = {ApJ},
         year = 2014,
        month = jun,
       volume = {787},
       number = {2},
          eid = {99},
        pages = {99},
          doi = {10.1088/0004-637X/787/2/99},
archivePrefix = {arXiv},
       eprint = {1404.1859},
 primaryClass = {astro-ph.GA},
       adsurl = {https://ui.adsabs.harvard.edu/abs/2014ApJ...787...99S}
}

@ARTICLE{Stil2025,
       author = {{Stil}, Jeroen M.},
        title = "{Rotation Measure Jitter: Wavelength-dependent Scatter in Rotation Measure Related to Faraday Complexity}",
      journal = {ApJ},
         year = 2025,
        month = jul,
       volume = {987},
       number = {2},
          eid = {173},
        pages = {173},
          doi = {10.3847/1538-4357/add69e},
archivePrefix = {arXiv},
       eprint = {2505.06460},
 primaryClass = {astro-ph.GA},
       adsurl = {https://ui.adsabs.harvard.edu/abs/2025ApJ...987..173S}
}

@ARTICLE{Taylor2009,
       author = {{Taylor}, A.~R. and {Stil}, J.~M. and {Sunstrum}, C.},
        title = "{A Rotation Measure Image of the Sky}",
      journal = {ApJ},
         year = 2009,
        month = sep,
       volume = {702},
       number = {2},
        pages = {1230-1236},
          doi = {10.1088/0004-637X/702/2/1230},
       adsurl = {https://ui.adsabs.harvard.edu/abs/2009ApJ...702.1230T}
}

@ARTICLE{Taylor2024,
       author = {{Taylor}, A.~R. and {Legodi}, L.~S.},
        title = "{A MeerKAT Polarization Survey of Southern Calibration Sources}",
      journal = {AJ},
         year = 2024,
        month = jun,
       volume = {167},
       number = {6},
          eid = {273},
        pages = {273},
          doi = {10.3847/1538-3881/ad4150},
archivePrefix = {arXiv},
       eprint = {2405.04131},
 primaryClass = {astro-ph.GA},
       adsurl = {https://ui.adsabs.harvard.edu/abs/2024AJ....167..273T}
}

@ARTICLE{Tjemsland2024,
       author = {{Tjemsland}, J. and {Meyer}, M. and {Vazza}, F.},
        title = "{Constraining the Astrophysical Origin of Intergalactic Magnetic Fields}",
      journal = {ApJ},
         year = 2024,
        month = mar,
       volume = {963},
       number = {2},
          eid = {135},
        pages = {135},
          doi = {10.3847/1538-4357/ad22dd},
archivePrefix = {arXiv},
       eprint = {2311.04273},
 primaryClass = {astro-ph.HE},
       adsurl = {https://ui.adsabs.harvard.edu/abs/2024ApJ...963..135T}
}

@INPROCEEDINGS{Vacca2015,
       author = {{Vacca}, V. and {Oppermann}, N. and {Ensslin}, T.~A. and {Selig}, M. and {Junklewitz}, H. and {Greiner}, M. and {Jasche}, J. and {Hales}, C.~A. and {Reneicke}, M. and {Carretti}, E. and {Feretti}, L. and {Ferrari}, C. and {Giovannini}, G. and {Govoni}, F. and {Horellou}, C. and {Ideguchi}, S. and {Johnston-Hollitt}, M. and {Murgia}, M. and {Paladino}, R. and {Pizzo}, R. and {Scaife}, A.},
        title = "{Statistical methods for the analysis of rotation measure grids in large scale structures in the SKA era}",
    booktitle = {Advancing Astrophysics with the Square Kilometre Array (AASKA14)},
         year = 2015,
        month = apr,
          eid = {114},
        pages = {114},
          doi = {10.22323/1.215.0114},
archivePrefix = {arXiv},
       eprint = {1501.00415},
 primaryClass = {astro-ph.CO},
       adsurl = {https://ui.adsabs.harvard.edu/abs/2015aska.confE.114V}
}

@ARTICLE{Vacca2016,
       author = {{Vacca}, V. and {Oppermann}, N. and {En{\ss}lin}, T. and {Jasche}, J. and {Selig}, M. and {Greiner}, M. and {Junklewitz}, H. and {Reinecke}, M. and {Br{\"u}ggen}, M. and {Carretti}, E. and {Feretti}, L. and {Ferrari}, C. and {Hales}, C.~A. and {Horellou}, C. and {Ideguchi}, S. and {Johnston-Hollitt}, M. and {Pizzo}, R.~F. and {R{\"o}ttgering}, H. and {Shimwell}, T.~W. and {Takahashi}, K.},
        title = "{Using rotation measure grids to detect cosmological magnetic fields: A Bayesian approach}",
      journal = {A\&A},
         year = 2016,
        month = jun,
       volume = {591},
          eid = {A13},
        pages = {A13},
          doi = {10.1051/0004-6361/201527291},
archivePrefix = {arXiv},
       eprint = {1509.00747},
 primaryClass = {astro-ph.CO},
       adsurl = {https://ui.adsabs.harvard.edu/abs/2016A&A...591A..13V}
}

@ARTICLE{Vacca2026,
       author = {{Vacca}, Valentina and {Hutschenreuter}, Sebastian and {Cabriolu}, Andrea and {Ensslin}, Torsten A. and {Roth}, Jakob and {Reineke}, Martin and {Frank}, Philipp and {Govoni}, Federica and {Murgia}, Matteo and {Fenu}, Gianni},
        title = "{DEFROST: Detecting Excess in Faraday Rotation thrOugh Sophisticated analysis Techniques}",
      journal = {arXiv e-prints},
         year = 2026,
        month = may,
          eid = {arXiv:2605.13605},
        pages = {arXiv:2605.13605},
          doi = {10.48550/arXiv.2605.13605},
archivePrefix = {arXiv},
       eprint = {2605.13605},
 primaryClass = {astro-ph.IM},
       adsurl = {https://ui.adsabs.harvard.edu/abs/2026arXiv260513605V}
}

@ARTICLE{VanEck2023,
       author = {{Van Eck}, C.~L. and {Gaensler}, B.~M. and {Hutschenreuter}, S. and {Livingston}, J. and {Ma}, Y.~K. and {Riseley}, C.~J. and {Thomson}, A.~J.~M. and {Adebahr}, B. and {Basu}, A. and {Birkinshaw}, M. and {En{\ss}lin}, T.~A. and {Heald}, G. and {Mao}, S.~A. and {McClure-Griffiths}, N.~M.},
        title = "{RMTable2023 and PolSpectra2023: Standards for Reporting Polarization and Faraday Rotation Measurements of Radio Sources}",
      journal = {ApJSS},
         year = 2023,
        month = aug,
       volume = {267},
       number = {2},
          eid = {28},
        pages = {28},
          doi = {10.3847/1538-4365/acda24},
archivePrefix = {arXiv},
       eprint = {2305.16607},
 primaryClass = {astro-ph.IM},
       adsurl = {https://ui.adsabs.harvard.edu/abs/2023ApJS..267...28V}
}

@ARTICLE{Vernstrom2023,
       author = {{Vernstrom}, Tessa and {West}, Jennifer and {Vazza}, Franco and {Wittor}, Denis and {Riseley}, Christopher John and {Heald}, George},
        title = "{Polarized accretion shocks from the cosmic web}",
      journal = {Science Advances},
         year = 2023,
        month = feb,
       volume = {9},
       number = {7},
          eid = {eade7233},
        pages = {eade7233},
          doi = {10.1126/sciadv.ade7233},
archivePrefix = {arXiv},
       eprint = {2302.08072},
 primaryClass = {astro-ph.CO},
       adsurl = {https://ui.adsabs.harvard.edu/abs/2023SciA....9E7233V}
}

\end{document}